\documentclass[a4paper]{article}
\usepackage{listings}
\usepackage{xcolor}
\usepackage{multirow}
\usepackage{jheppub}  
\usepackage{dcolumn}
\usepackage[english]{babel}
\usepackage[utf8x]{inputenc}
\usepackage[T1]{fontenc}
\usepackage{palatino}
\usepackage{amsmath}
\usepackage{afterpage}
\usepackage{graphicx, subcaption}
\usepackage[colorinlistoftodos]{todonotes}
\usepackage[colorlinks=true]{hyperref}
\usepackage{makeidx}
\newcommand{\be}{\begin{equation}}  
\newcommand{\ee}{\end{equation}}  
\newcommand{\bea}{\begin{eqnarray}}  
\newcommand{\eea}{\end{eqnarray}}  
\makeindex
\begin{document}

\vspace*{1.2cm}

\thispagestyle{empty}
\begin{center}
{\LARGE \bf Explicitly exotic heavy flavor mesons}

\par\vspace*{7mm}\par

{

\bigskip

\large \bf K. P. Khemchandani$^{1,*}$, Taísa Veloso$^1$, G. Peres de Andrade$^2$, Hao-Nan Liu$^2$, A.~Martínez Torres$^2$, Luciano M. Abreu$^3$, Philipp Gubler$^4$}

\bigskip

{\large \bf  E-Mail$^*$: kanchan.khemchandani@unifesp.br}

\bigskip

{$^1$Universidade Federal de S\~ao Paulo, C.P. 01302-907, S\~ao Paulo, Brazil.
\\
$^2$Instituto de F\'{\i}sica, Universidade de S\~{a}o Paulo, Rua do Mat\~{a}o, São Paulo SP, 05508-090, Brazil.\\
Instituto de F\'isica, Universidade Federal da Bahia, 40210-340, Salvador, BA, Brazil.\\
Advanced Science Research Center, Japan Atomic Energy Agency, Tokai, Ibaraki 319-1195, Japan.}

\bigskip

{\it Presented at the Workshop of Advances in QCD at the LHC and the EIC, CBPF, Rio de Janeiro, Brazil, November 9-15 2025}


\vspace*{15mm}

\end{center}
\vspace*{1mm}

\begin{abstract}

In this talk, I summarize the highlights of our recent and ongoing work on exotic hadrons involving explicitly exotic heavy-flavor mesons. By heavy flavors, I refer in particular to strange and charm quarks. As discussed in this manuscript, several puzzle-like features appearing in recent experimental data can be understood in terms of coupled-channel interactions. The main result is the prediction of an exotic meson carrying both charm and strangeness quantum numbers, whose properties cannot be explained as those of a simple quark–antiquark bound state. I also discuss the future prospects of our research program on this topic. 
\end{abstract}

 
 \section{Introduction}
The existence of exotic hadrons in nature has been firmly established through numerous experimental observations reported over the past two decades.~(for some recent reviews, see \cite{Zou:2026ngp,Wang:2025ujb,Shen:2025jjm}). Even though there is a long list of exotic hadrons that have now been discovered, the challenge to understand their structure remains to be handled and requires dedicated and consistent investigations. It is worth citing some examples of interesting observations that can be made from experimental data obtained for hadronic systems which would naively appear to be uninteresting. For example, Ref.~\cite{Zhang:2024qwv} provides an invariant mass spectrum for the $J\Psi \phi$ system. Interestingly, this mass spectrum shows several peak structures, which have been identified with states such as $\chi_{c1}(4140)$, $\chi_{c1}(4274)$, $\chi_{c1}(4500)$, etc. A first thought about the $J\Psi \phi$ system that could come to our mind could be that the system could not show any interesting physics since the $J\Psi \phi$ interactions are expected to be OZI-suppressed (Okubo-Zweig-Izuka rule).  In fact, we would not be wrong to consider such a thought. The key, however, is not to stop at this thought, elaborate a bit more and consider coupled channel interactions, with channels such as $D_s^{(*)}\bar D_s^{(*)}$. It is possible to consider an exchange of light mesons in these latter channels, as they contain one light quark, and then the interactions are no more suppressed. We find it useful to cite our former work on the $D_s^{(*)}\bar D_s^{(*)}$ system~\cite{MartinezTorres:2016cqv}, where we show that the interaction in such a system can lead to the formation of some molecular states. It is worth citing also a very detailed work done, using an alternative formalism, in Ref.~\cite{Molina:2009ct}.

Another interesting system, which, on the face, would look like related to OZI-suppressed interactions, and which led to some debates in recent times, is that of the $\phi$-nucleon system. The enigma started with the publication of the correlation function by the Alice collaboration in Ref.~\cite{ALICE:2021cpv}, which shows that the spin-averaged interactions must be attractive in nature. Providing a quick definition of the correlation function can be useful for young readers here. Correlation functions are a new type of observable in hadron physics, which are bringing useful information on the interactions in a particular system. The experimentalists measure a normalized ratio  of a number of pairs of a given hadronic system, as a function of their relative momenta, coming from the same event versus a product of the number of same hadrons measured separately as coming from separate events. Such a ratio, if divergent from unity, can indicate the presence of a resonance/bound state formed in the system, depending on the divergence being less than or more than unity at small relative momentum. Interestingly, the  scattering length for the $\phi N$ interaction in total spin-3/2 has also been determined by lattice QCD simulations~\cite{Lyu:2022imf} with the value being  $a^{(3/2)}=-1.43(23)^{+36}_{-06}$ fm. The aforementioned value, once again, indicates an attractive $\phi N$ interaction in the spin 3/2 case, at least. In contrast, the scattering length determined from the photoproduction of the $\phi$-meson indicates weak interactions~\cite{Strakovsky:2020uqs}. As we will show in more detail in the next section, we had already predicted that a resonance like structure appears in the spin 3/2 configuration of the $\phi N$ system about a decade ago~\cite{Khemchandani:2011et}. We have shown that such a structure appears due to coupled channel interaction, which breaks the apparent OZI suppressed picture of the system and can describe the experimental data on the correlation function using our previous results \cite{Khemchandani:2011et,Khemchandani:2011mf} obtained within the formalism developed in Ref.~\cite{Khemchandani:2011et,Khemchandani:2013nma}.

Finally, another recent work, which we would like to discuss here, is that which predicted the existence of an exotic isovector state with explicit charm and strange content~\cite{Malabarba:2022pdo}. This work was motivated by the first experimental observation of two exotic states, $T_{\bar c\bar s}(2870)$ and $T_{\bar c\bar s}(2900)$, by the LHCb collaboration in the $D^-K^+$ invariant mass distribution, in Refs.~\cite{LHCb:2020pxc,LHCb:2020pxc}. The same states have recently been confirmed  in a different production channel~\cite{LHCb:2024vfz}. Both the states would require a minimum quark content $\bar cd\bar s u$, thus being explicitly exotic in nature. Their spin parities are different:  $T_{\bar c\bar s}(2870)$  is a $0^+$ state, while $T_{\bar c\bar s}(2900)$ is a vector, $1^-$ state. According to the recent publication of LHCb~\cite{LHCb:2024vfz}, both states seem to have a width of around 100 MeV. Several works, followed  by the proposal made in Refs.~\cite{Liu:2020nil,Huang:2020ptc,Molina:2020hde}, agree on a $\bar D^*K^*$ moleculelike description for the scalar  $T_{\bar c\bar s}(2870)$ state. Providing a similar description for the vector state, as generated by s-wave two-meson interactions, is a difficult task when considering pseudoscalar/vector nonets. However, an interesting study was made by the authors of Refs.~\cite{Dong:2020rgs,He:2020btl,Qi:2021iyv,Chen:2021tad} who considered a system comprising an axial meson and a pseudoscalar meson to describe $T_{\bar c\bar s}(2900)$. These former works considered either $\bar D_1(2420)$K or $\bar DK_1(1270)$ system. Ideally, one should couple the two former systems to draw more robust conclusions.  With the idea of making a more complete study, and keeping in mind that $D_1(2420)$ can be understood as a state generated from $D\rho$ coupled channel interactions (as shown in Ref.~\cite{ Gamermann:2007fi}), together with the possibility of generating $K_1(1270)$ in $K\rho$ and coupled channel systems (following Refs.~\cite{Geng:2006yb,Roca:2005nm,}), we studied a three body system: $K \rho \bar D$. To make such a study, we solve the Faddeev equations in the fixed center approximation. The inputs required to solve these equations are two body $t$-matrices, which are first obtained by solving coupled channel Bethe-Salpeter equations. In this way, we incorporate $D_1(2420)K$ as well as $DK_1(1270)$ in the same study. As we discuss in the subsequent section, we do not succeed in describing $T_{\bar c\bar s}(2900)$. Our finding implies that a three-body treatment, which considers full two-body interactions and not just the resonant part, is different than studying $D_1(2420)K$ and $DK_1(1270)$ separately. In the same work~\cite{Malabarba:2022pdo}, we investigated the system with opposite charm too, $K\rho D$. Here, we have an additional strongly attractive subsystem, $KD$. In this case, we end up finding the generation of an isoscalar as well as an isovector state. We identify the isoscalar state with $D^*_{s1}(2860)$, and the exotic isoscalar state necessarily would require a minimum quark content of $c \bar d\bar s u$. Clearly, if states with minimum quark content of $\bar c  d\bar s u$ have been observed, we should soon be able to find a state with opposite charm as found in our work. We are now studying further properties of both isoscalar and isovector states  found in Ref.~\cite{Malabarba:2022pdo}.

\section{Mathematical Framework and  a summary of findings}

The main objective of this talk is to discuss the explicitly exotic heavy flavor meson found in our work~\cite{Malabarba:2022pdo}. Before proceeding with the discussions of the formalism, it is important to recall the importance of such exotic states. The existence of such states is not only interesting from the point of view of understanding the working of strong interactions at low energies, but their presence can be important in understanding what would otherwise look like unresolved puzzles. A useful example to present would be our work on the studies of the interaction of open charm mesons, which could produce $\chi_{c1}(3872)$ together with pions and rhos. We also studied the inverse processes, considering that a large number of light mesons would be present in heavy-ion collisions after the hadronization phase. In Ref.~\cite{MartinezTorres:2014son,Abreu:2016qci} we investigated how the yield of exotic states is affected by the production and absorption of $\chi_{c1}(3872)$ through the aforementioned processes. In yet another study~\cite{Khemchandani:2016ftn}, we discussed that there exists a disagreement between the measured ratio of abundance of $\Xi^-$  to that of $\Lambda, \Sigma^0$, when compared with the results of the statistical model and the relativistic transport code. We pointed out that strong $\bar K \Sigma$, $\bar K^*\Sigma$, and similar two-body interactions can affect such an abundance. Hence, understanding the nature and properties of exotics is relevant for several purposes. 

\begin{figure}[h!]
\begin{center}
\includegraphics[width=0.94\textwidth]{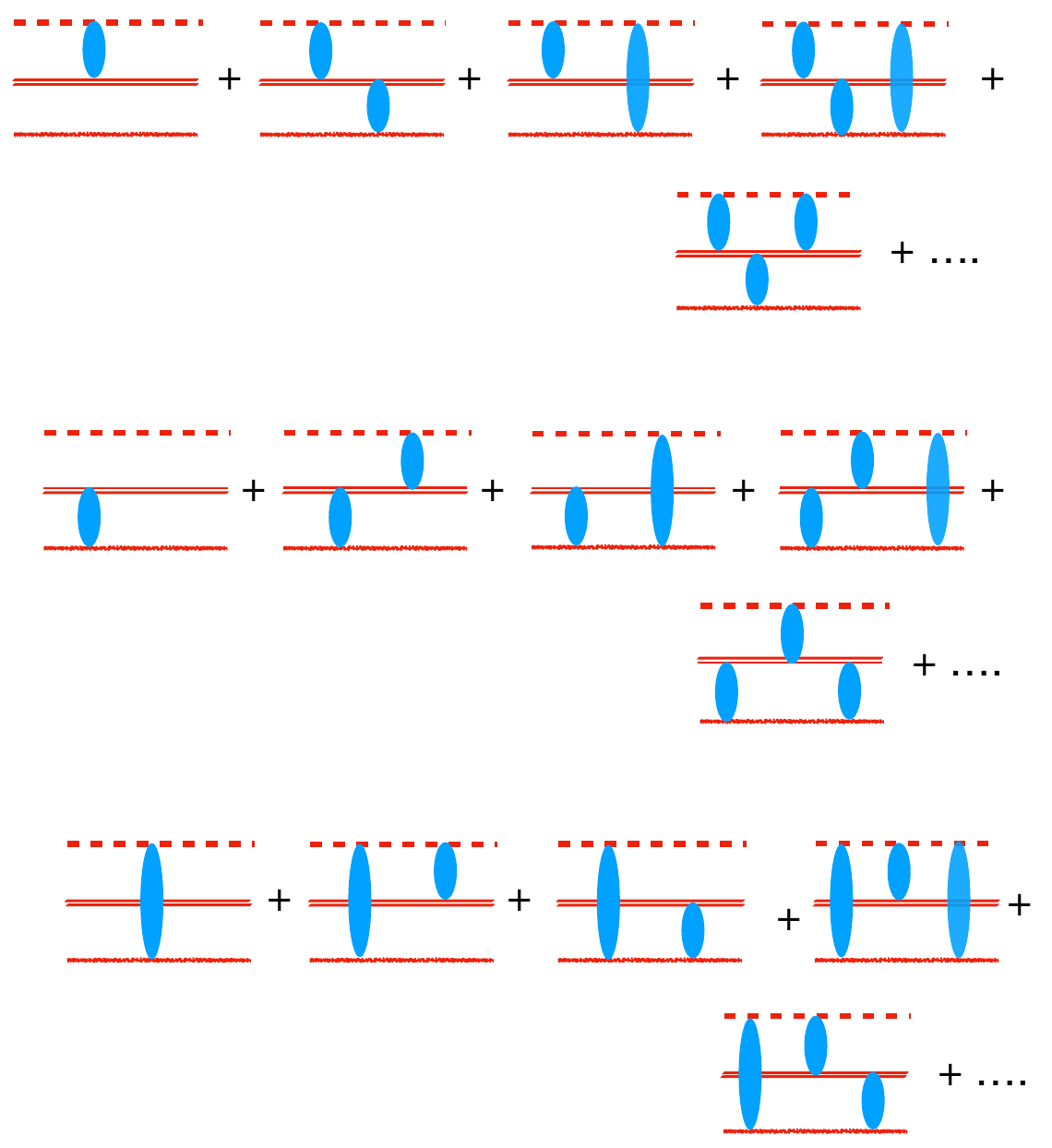}
\caption{Diagrams for three coupled series constituting  Faddeev equations, which consider all different permutations of interactions among different pairs of a three-body system. The dashed lines represent a kaon, double lines represent $\rho$ and thick lines represent $D$.}
\label{fig1}
\end{center}
\end{figure}

As mentioned in the introduction, we studied $K\rho \bar D$ to test the possibility of describing the $T_{\bar c\bar s}(2900)$ state in terms of three-body dynamics where $\bar D_1 \rho$ could resonate as $D_1(2420)$ and the $K\rho$ state could form $K_1(1270)$ simultaneously. In this way, we could treat $\bar D_1(2420) \rho$ and $\bar D K_1(1270)$ as coupled channels, which were studied separately and where the formation of $T_{\bar c\bar s}(2900)$ was reported in Refs.~\cite{Dong:2020rgs,He:2020btl,Qi:2021iyv,Chen:2021tad}. The problem of treating $\bar D_1(2420) \rho$ and $\bar D K_1(1270)$ separately is not only the missing coupling between the two channels. One could also question the missing $D_1(2430) \rho$ interactions in the background, considering the similarity of the masses of the two $D_1$ states (for more discussions on the two states, see Ref.~\cite{Khemchandani:2023xup}).  Thus, studying  $K\rho \bar D$ interactions can be considered as a more complete treatment. To study the three-body interaction, one needs to consider three different series of interactions among the subsystems. An illustration of such a series is presented in Fig.~\ref{fig1}, which constitutes the Faddeev equations
\begin{align}
    T^1&=t^1+t^1G\left(T^2+T^3\right),\\\nonumber
       T^2&=t^2+t^2G\left(T^1+T^3\right),\\\nonumber
          T^3&=t^3+t^3G\left(T^1+T^2\right).
\end{align}
The input $t^i$, in the set of equations stated above, implies that the particle $i$ is a spectator in the diagram and represents $t$-matrices which are obtained by solving nonperturbative two-body scattering equations
\begin{align}\label{bseq}
t=v+vgt,
\end{align}
onde $v$ is the lowest-order two-body interaction which enters as kernel to solve the scattering equations, and $g$ is a two-body Green's function.

\begin{figure}[h!]
     \centering
     \includegraphics[width=0.8\linewidth]{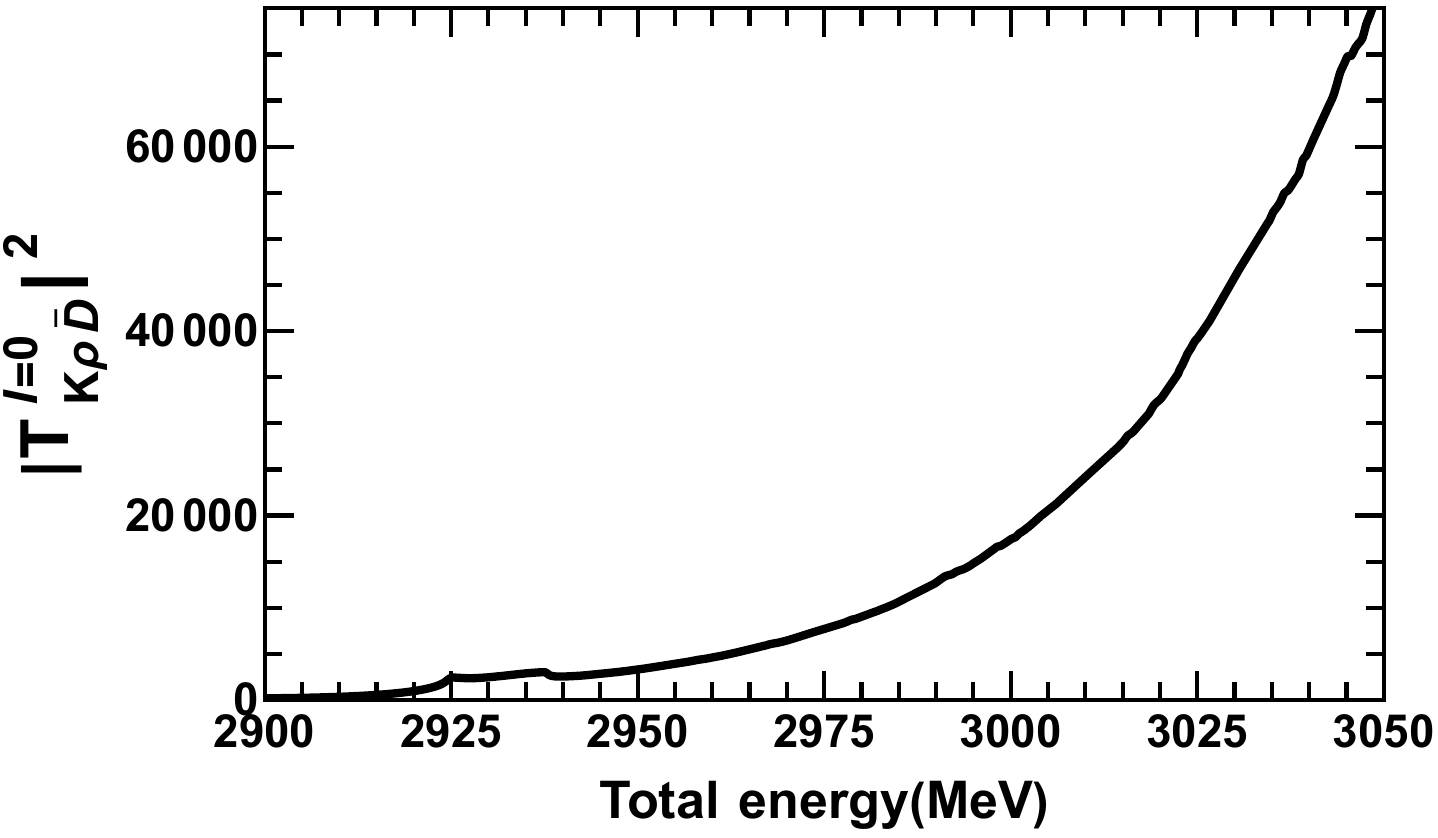}
     \caption{Squared three-body amplitude for $K\rho\bar D$ system having  total isospin 0.}
     \label{fig2}
 \end{figure}

 Solving Faddeev equations exactly is an extremely complex procedure, and certain approximations suitable for different types of three-body systems are often considered. For example, in the case of $K\rho\bar D$, the two-body $\rho\bar D$ forms a quasibound state $\bar D_1(2420)$, which lies more than 100 MeV below the threshold of the $D\rho$ system. In such a case, a suitable approximation is considering that the $D\rho$ pair remains unperturbed during the interaction with the third particle, which is a kaon. Such an approximation is frequently referred to as the fixed center approximation (FCA)~\cite{MartinezTorres:2020hus}. To solve the Faddeev equations within FCA, we determine the $K\rho$ and $K\bar D$ $t$-matrices. For the former case, we follow Ref.~\cite{Geng:2006yb} and include $\pi K^*$, $\rho K$, $\phi K$, $\omega K$ and  $K^*\eta$ as coupled channels. It has been shown in Ref.~\cite{Geng:2006yb} that the aforementioned coupled interaction generates a resonance which can be well identified with $K_1(1270)$. The same work shows a good description of the experimental data on $K^-p\to K^-\pi^+\pi^-p$ obtained with the amplitudes determined in Ref.~\cite{Geng:2006yb}, demonstrating the reliability of the results of the work. The kaon-$\bar D$ $t$-matrix was obtained following Ref.~\cite{Guo:2006fu}, where a Lagrangian based on heavy quark and chiral symmetries is used, recalling that both heavy and light quarks are present in the system. The interaction in this system is weakly attractive and does not lead to the formation of a state. On solving three-body equations we do not find the presence of any state (see Fig.~\ref{fig2}) and, hence, we do not find a description for $T_{\bar c\bar s}(2900)$ discovered by the LHCb collaboration~\cite{LHCb:2020pxc,LHCb:2020pxc,LHCb:2024vfz}, which implies that $T_{\bar c\bar s}(2900)$ might have a different nature.
It is useful to point out here that our work is different than those presented in Refs.~\cite{Dong:2020rgs,He:2020btl,Qi:2021iyv,Chen:2021tad}, where $\bar D_1(2420)K$ and $\bar DK_1(1270)$ interactions have been found to generate a resonance. In our work, $D\rho/~ K\rho$ interactions are present at the same time. Besides, the coupled channel interaction used in our work does not have the information on $D_1(2420)/K_1(1270)$ alone, our amplitudes contain a background from detailed coupled channel interactions, including having a finite width for $D_1(2420)/K_1(1270)$. 

\begin{figure}[h!]
     \centering
     \includegraphics[width=0.45\linewidth]{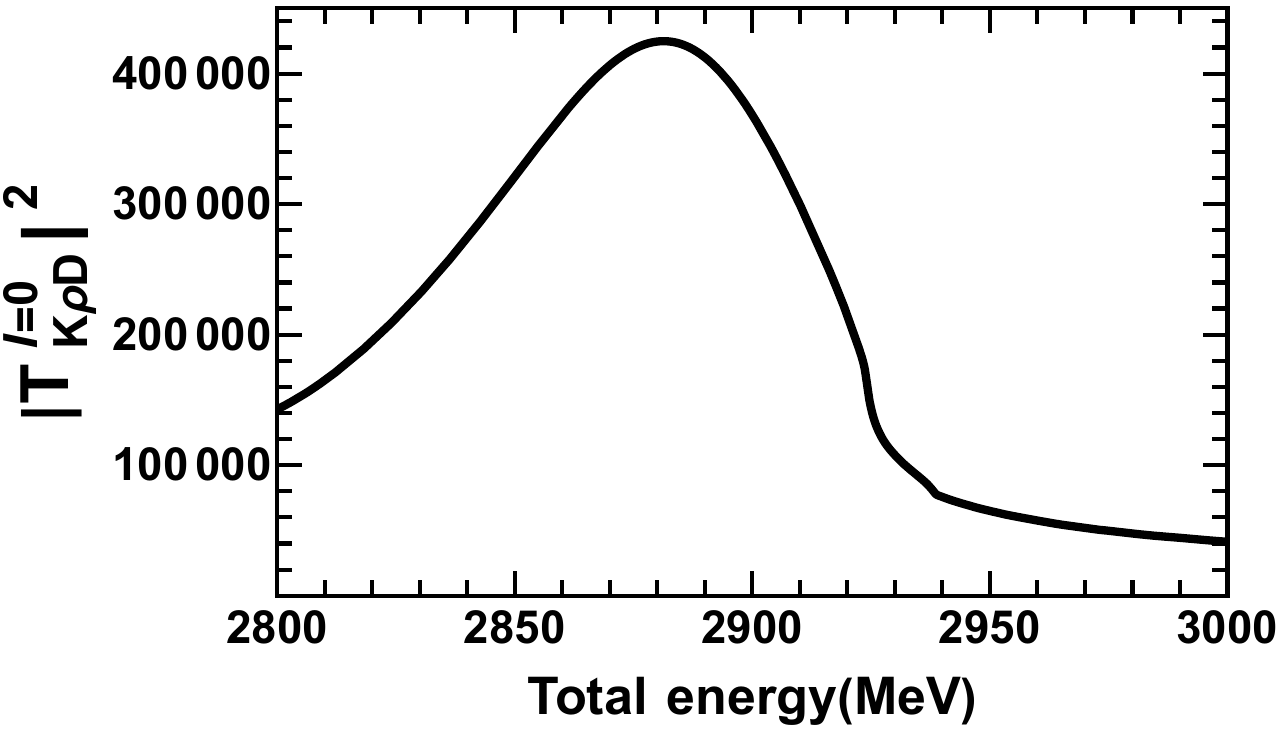}     \includegraphics[width=0.44\linewidth]{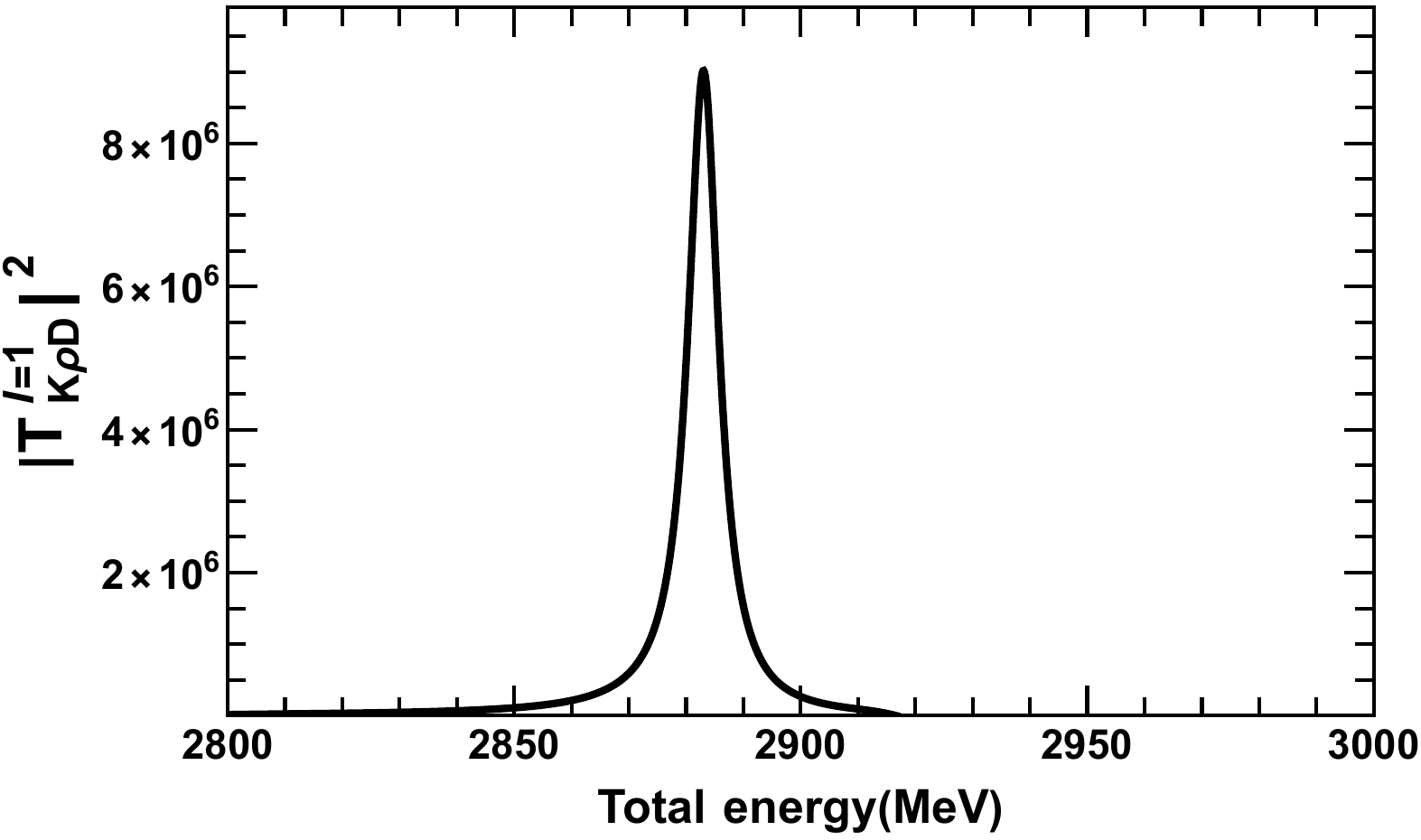}
     \caption{Squared three-body amplitude for $K\rho D$ system having  total isospin 0 (left) and 1 (right).}
     \label{fig3and4}
 \end{figure}

We found it useful to study the system made by the same mesons except for inverting the charm quantum number. An important difference is that the $DK$ interaction is strongly attractive in this case and forms $D_{s0}(2317)$. The replacement of weak $\bar DK$ interaction by $DK$, leads to the formation of a state in total isospin 0 as well as in isospin 1 (as shown in Fig.~\ref{fig3and4}).
 We relate the state found in the isospin 0 case with $D^*_{s1}(2860)$. Very little information is available on this state, and our work can be useful in better understanding its properties. The state found in the isovector case is completely exotic and cannot be described in the traditional quark model. This state can be interpreted as an effective $D^*_{s0}(2317)\rho$ resonance. 

 \section{Summary and outlook}
It has been pointed out in the present talk that, considering the presence of exotic 
 hadrons can be very important in understanding some puzzling experimental data published in recent times. The highlight of this talk is the finding of three-body resonances arising from $K\rho D$ dynamics, out of which the isoscalar state can be identified as $D^*_{s1}(2860)$. A narrow  exotic state with isospin 1 has also been found, which we hope can be confirmed in upcoming experimental studies. Since reconstructing three-body invariant masses can be more challenging, in order to encourage experimental searches of the exotic state found in our work, we are currently studying its decay to different two-body systems. A similar investigation is also being carried out to better determine the properties of $D^*_{s1}(2860)$.
 
\section*{Acknowledgements}
The authors thank CNPq (L.M.A.: Grant No. 309950/2020-1, 400215/2022-5, 200567/2022-5; K.P.K: Grants No. 407437/ 2023-1 and No. 306461/2023-4; A.M.T: Grant No. 304510/2023-8), FAPESP (K.P.K.: Grant Number 2022/08347-9; A.M.T.: Grant number 2023/01182-7); and CNPq/FAPERJ under the Project INCT-F\'{\i}sica Nuclear e Aplica\c c\~oes (Contract No. 464898/2014-5).

\end{document}